\documentclass[11pt]{article}

\usepackage[letterpaper,margin=1in]{geometry}
\usepackage[T1]{fontenc}
\usepackage[utf8]{inputenc}
\usepackage{lmodern}
\usepackage{microtype}
\usepackage{graphicx}
\usepackage{amsmath,amssymb,mathtools}
\usepackage{amsthm}
\usepackage{booktabs}
\usepackage{multirow}
\usepackage{tabularx}
\usepackage{array} 
\usepackage{xcolor}
\usepackage{listings}
\usepackage{pifont}
\usepackage[numbers,sort&compress]{natbib}
\usepackage[hidelinks]{hyperref}

\graphicspath{{Fig/}{_legacy/Fig/}}

\theoremstyle{plain}

\theoremstyle{definition}

\newcommand{\nmcp}{\textsc{nuance-mcp}}
\newcommand{\adapter}{\texttt{MicroscopeAdapter}}
\renewcommand{\cap}[1]{\texttt{#1}}   
\newcommand{\tool}[1]{\texttt{\nolinkurl{#1}}}    

\newcommand{\temagent}{TEM\,Agent}
\newcommand{\cmark}{\textcolor{green!60!black}{\ding{51}}}
\newcommand{\xmark}{\textcolor{red!70!black}{\ding{55}}}

\begin{document}

\title{Schema-Bound LLM Control of Scientific Instrumentation through Model Context Protocol Skills}

\author{%
Roberto dos Reis\textsuperscript{1,2,*}
\and
Vinayak P.~Dravid\textsuperscript{1,2}
}
\date{}

\hypersetup{
  pdftitle={Schema-Bound LLM Control of Scientific Instrumentation through
            Model Context Protocol Skills},
  pdfauthor={Roberto dos Reis, Vinayak P. Dravid},
  pdfkeywords={Model Context Protocol; MCP skills; large language models;
               agentic instrumentation; schema-bound tool typing;
               electron microscopy; tool calling; provenance},
  pdfsubject={Agentic control protocols for scientific instrumentation},
}
\maketitle
\begin{center}
\small
\textsuperscript{1}Department of Materials Science and Engineering, Northwestern University, Evanston, IL 60208, USA\\
\textsuperscript{2}NU\textit{ANCE} Center, Northwestern University, Evanston, IL 60208, USA\\
\textsuperscript{*}Correspondence: \href{mailto:roberto.reis@northwestern.edu}{roberto.reis@northwestern.edu}\\
\end{center}

\begin{abstract}
Large language models can plan and execute tool-mediated scientific work, but scientific instruments remain hard to connect to such agents. Vendor application programming interfaces (APIs) load only inside the acquisition host process, facility policy discourages cloud-hosted agents, and natural-language interfaces routinely emit physically unreasonable arguments. In this work, we describe a \emph{method} for connecting local large language models (LLMs) to scientific instruments through the Model Context Protocol (MCP). The method has four components. First, a schema-bound tool surface validates every request against physical bounds before any vendor call is dispatched. Second, a host-process adapter pattern, defined by a vendor-neutral abstract base class, separates language-side reasoning from instrument-side execution. Third, a persistent live-processing job lifecycle promotes long-running analyses to first-class typed tools. Fourth, a \emph{skill} abstraction, implemented through MCP prompts that register parameterized tool sequences, composes typed tools into reusable multi-step protocols. As a reference implementation and software-only validation study, we provide an open-source server, \nmcp{}, that exposes 30 typed tools, 5 live-job types, and 6 skills, against a physics-plausible simulator that implements the same protocol surface. All validation reported here is software-only: every claim is demonstrated against the simulator, and live-instrument validation on commercial scientific hardware is intentionally out of scope and reported separately under the corresponding intellectual-property disclosure. All 120 hardware-independent tests (schema validation, simulator I/O, tool dispatch, live-job lifecycle, and bridge protocol) pass deterministically; the 15 local-LLM integration tests pass 12--15 of 15 across runs depending on model non-determinism. On identical hardware, a single-run, undersampled probe across five openly available tool-calling LLMs is consistent with the schema-bound interface being drivable by small open-weight models locally, without cloud dependency; the probe is preliminary, is not accompanied by confidence intervals, and is not a benchmark. Schema-bound tool typing, an adapter contract, a live-job lifecycle, and the skill abstraction together define a bounded, testable, and reproducible foundation for closed-loop agentic instrumentation research.
\end{abstract}


\section{Introduction}\label{sec:intro}
A scientific instrument is a programmable physical system, but the programming interface is rarely uniform. Each vendor exposes a Python or scripting surface that runs inside its own acquisition application, with idiosyncratic data types, threading conventions, and update cycles. The result is that day-to-day experimentation is mediated through graphical interfaces and bespoke scripts; reproducibility and language-first automation both suffer. Large language models (LLMs) are now competent planners for tool-mediated scientific work. Co-Scientist (Boiko et~al., 2023) showed that an LLM agent can plan and execute organic synthesis on cloud-lab hardware~\citep{Boiko2023Coscientist}; ChemCrow demonstrated that curated chemistry tools improve task fidelity and reduce hallucination~\citep{Bran2024ChemCrow}; self-driving laboratories now compose these ideas at the bench~\citep{Tom2024SDL}. The Berkeley Function-Calling Leaderboard~\citep{Patil2025BFCL} has made tool-calling itself a measurable skill, and modern open-weight models exceed 0.85 on its agentic split.

In microscopy, recent work has driven imaging and probe placement with active learning and hypothesis-driven agents~\citep{Kalinin2021AutoEM,Ziatdinov2022AtomAI,liu2023autonomous,Liu2025RewardSPM,Pratiush2025SEEK}, and dedicated LLM-agent platforms have appeared for atomic-force microscopy~\citep{Mandal2025AILA} and transmission electron microscopy~\citep{Wall2025TEMAgent,Chen2026EMSeek}. More recently,~\citet{Jamali2026ThinkingMicroscopes} framed the microscope itself as a thinking system that iteratively refines its own protocols. The most recent generation of agentic-science systems has widened this context beyond single-task tool use. Co-Scientist (Gottweis et~al., 2026), a system distinct from the earlier Coscientist of Boiko et~al.\ despite the near-identical name, uses a multi-agent architecture to generate, debate, rank, and refine testable biomedical hypotheses~\citep{Gottweis2026CoScientist}; Robin integrates literature agents with experimental-design and data-analysis agents in an iterative lab-in-the-loop discovery workflow~\citep{Ghareeb2026Robin}; and Empirical Research Assistance combines LLM code generation with tree search to produce expert-level empirical software for scorable scientific tasks~\citep{Aygun2026ERA}. These systems plan, generate executable artefacts, interpret measurements, and update their hypotheses. For physical experiments, that shift makes the underlying control layer more important, not less: agents that act on microscopes must reason about latency, calibration, noise, drift, damage thresholds, transfer functions, and the uncertainty that propagates from one measurement to the next.

\begin{figure}
  \centering
  \includegraphics[width=\columnwidth]{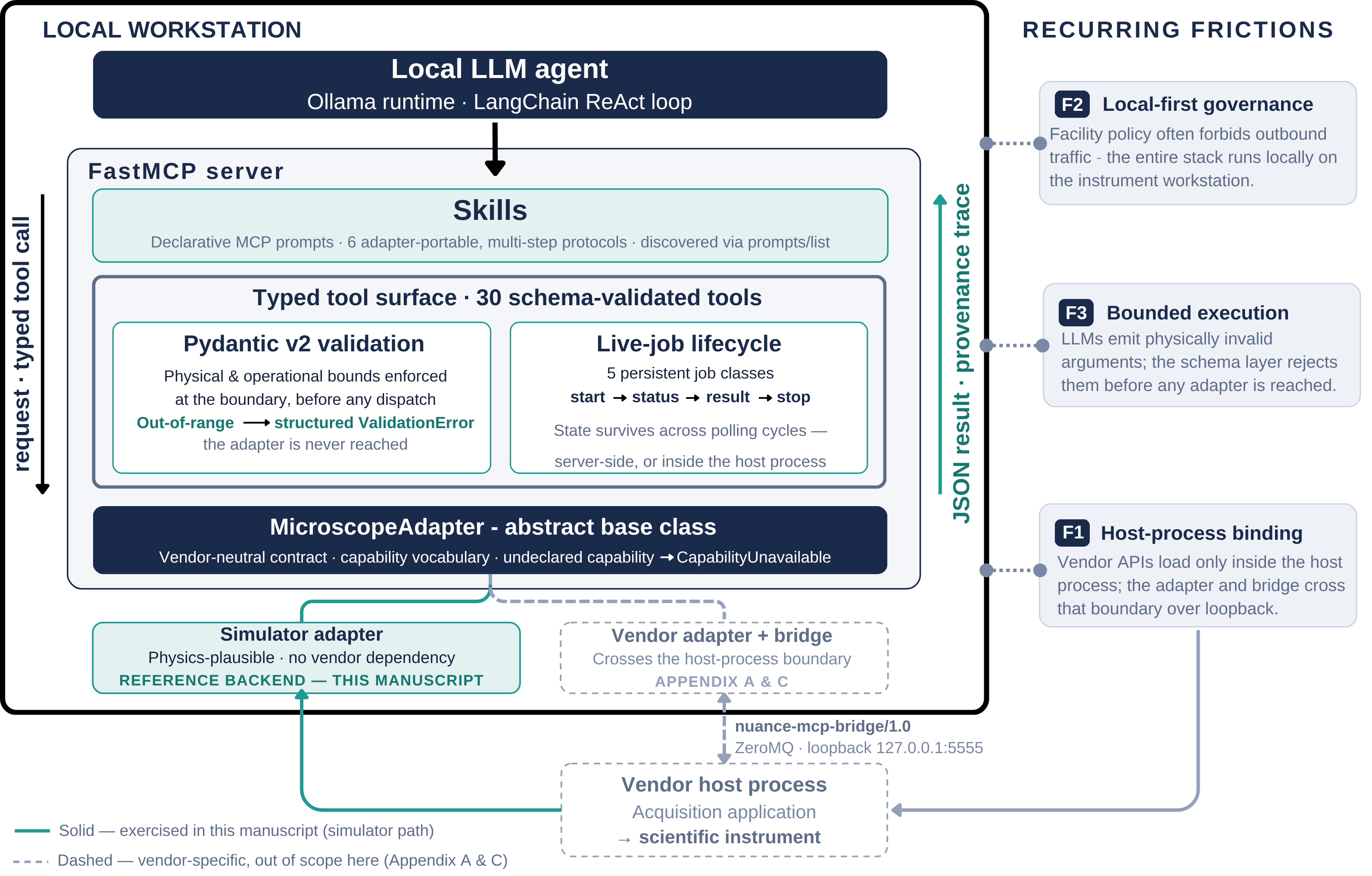}
 \caption{\textbf{MCP instrumentation control architecture.} The Model Context Protocol (MCP) connects a local LLM agent to a scientific instrument through a layered protocol surface that runs entirely on the instrument workstation. Three protocol layers sit above the backend: a \emph{skills} layer of declarative MCP prompts (Section~\ref{sec:skills}); a \emph{typed tool surface} of 30 schema-validated tools, within which Pydantic~v2 validation (Section~\ref{sec:schema}) and the persistent live-processing job lifecycle (Section~\ref{sec:live}) operate; and the vendor-neutral \texttt{MicroscopeAdapter} abstract base class (Section~\ref{sec:adapter}). Concrete adapters implement that contract: the physics-plausible simulator exercised throughout this manuscript, or a vendor adapter that crosses the host-process boundary through the versioned bridge (dashed; see Appendix~\ref{app:bridge} and Appendix~\ref{app:vendors}). The three recurring frictions of Section~\ref{sec:intro} are resolved at the layers indicated: host-process binding at the adapter contract and bridge, local-first governance by the workstation-local stack, and bounded execution at the schema layer.}
 \label{fig:architecture}
 \end{figure}

The Model Context Protocol (MCP), introduced by Anthropic in late 2024 and adopted across major model providers by 2026~\citep{Anthropic2024MCP,Hou2025MCPSurvey}, provides a standardised way to expose tools, resources, and prompts to any compatible model. We build on MCP, but adopting it for scientific instrumentation reveals three recurring frictions that the protocol itself does not resolve (Fig.~\ref{fig:architecture}).

\noindent\textbf{1: Host-process binding.} Most vendor APIs load only inside the running acquisition process. An external Python interpreter cannot import the vendor's bindings, so an agent that lives off the instrument PC cannot reach the column without crossing a process boundary.

\noindent\textbf{2: Local-first governance.} Many academic and industrial facilities forbid outbound connections from the instrument PC. Any cloud-only LLM backend is therefore off the table for routine use.

\noindent\textbf{3: Bounded execution at the tool boundary.} LLMs reliably emit syntactically valid tool calls with physically unreasonable arguments: a stage tilt of $95^\circ$, a negative exposure, a region of interest outside the image. A control
protocol must reject such arguments at the boundary, with auditable failure paths that do not depend on prompt phrasing. Reproducible, audit-friendly agent traces are now an active research direction~\citep{LLMProvenance2025,RLAM2026,AuditTrails2026}.

We argue that these are protocol-design problems rather than instrument-engineering problems. Once the tool contract, the host-process boundary, the live-job lifecycle, and the skill abstractions are right, the specific instrument software becomes a swap-in component. We describe a method that combines (i) a schema-bound MCP tool surface validated by Pydantic v2 models \citep{Colvin2024Pydantic} before dispatch; (ii) a vendor-neutral \adapter{} abstract base class that defines the surface every adapter implements, paired with a physics-plausible simulator and an optional host-process bridge for vendor APIs that require one; (iii) a persistent live-processing job lifecycle exposed as ordinary tools; and (iv) a skill abstraction, implemented through MCP prompts, that composes typed tools into reusable multi-step protocols.

This manuscript presents the method and its reference implementation as a \emph{software-only validation study}: every claim is demonstrated against a physics-plausible simulator that implements the same \adapter{} contract, and the test suite, simulator, and source are released under MIT license (Section~\ref{sec:results}). Live-instrument validation on commercial scientific hardware is intentionally out of scope here and is reported separately under the corresponding intellectual-property disclosure \citep{dosReis2025Disclosure}; readers should therefore note that on-hardware performance cannot be assessed from this manuscript alone.

The protocol-level framing distinguishes this work from the closest prior art. \temagent{} \citep{Wall2025TEMAgent} demonstrated that MCP-mediated TEM workflows are feasible by exposing a vendor-specific surface to an LLM agent. Our contribution sits at a different layer of the stack. First, schema-bound validation runs in the server \emph{before} dispatch to any adapter, so out-of-range or malformed requests fail at the boundary with a structured error and never reach the instrument. Second, the host-process boundary is crossed by a minimal, versioned JSON contract over loopback rather than by embedding the agent in the vendor process, which keeps the LLM stack swappable and the bridge independently testable. Third, a paired physics-plausible simulator shares the schema layer with any live adapter, so the entire validation suite runs without instrument occupancy. Fourth, a persistent live-processing job lifecycle and a skill abstraction - registered as MCP prompts so they are discoverable through ordinary \verb|prompts/list| calls - compose typed tools into multi-step procedures. Together these choices yield a protocol-design contribution rather than an autonomy demonstration: a bounded, testable, local foundation for closed-loop and autonomous microscopy research.

The reference implementation, \nmcp{}, is open source. It ships with a physics-plausible simulator adapter that exercises the full protocol surface. The remainder of this paper is organised as follows. Section~\ref{sec:methods} presents the four design choices that define the method. Section~\ref{sec:results} reports the software validation outcomes against the simulator and the local-LLM probe on the same hardware. Section~\ref{sec:discussion} situates the contribution in context and states its limits. Appendix~\ref{app:bridge} documents the versioned JSON bridge contract that vendor-specific adapters must implement when an instrument API is host-process bound. Appendix~\ref{app:vendors} gives implementation guidance for adapting the contract to vendor-specific platforms.
\section{Method}\label{sec:methods}
The method consists of four coupled design choices. The first two address frictions identified in Section~\ref{sec:intro}; the deployment model (Section~\ref{sec:deploy}) addresses local-first governance; the third and fourth introduce abstractions (a live-job lifecycle exposed as typed tools, and a skill abstraction implemented through MCP prompts) that compose into multi-step protocols. Apart from the (vendor-specific) bridge layer, every layer is shared across adapter implementations and is exercised end-to-end against the reference simulator.
\subsection{Schema-bound tool surface}\label{sec:schema}
\begin{figure*}
  \centering
  \includegraphics[width=\textwidth]{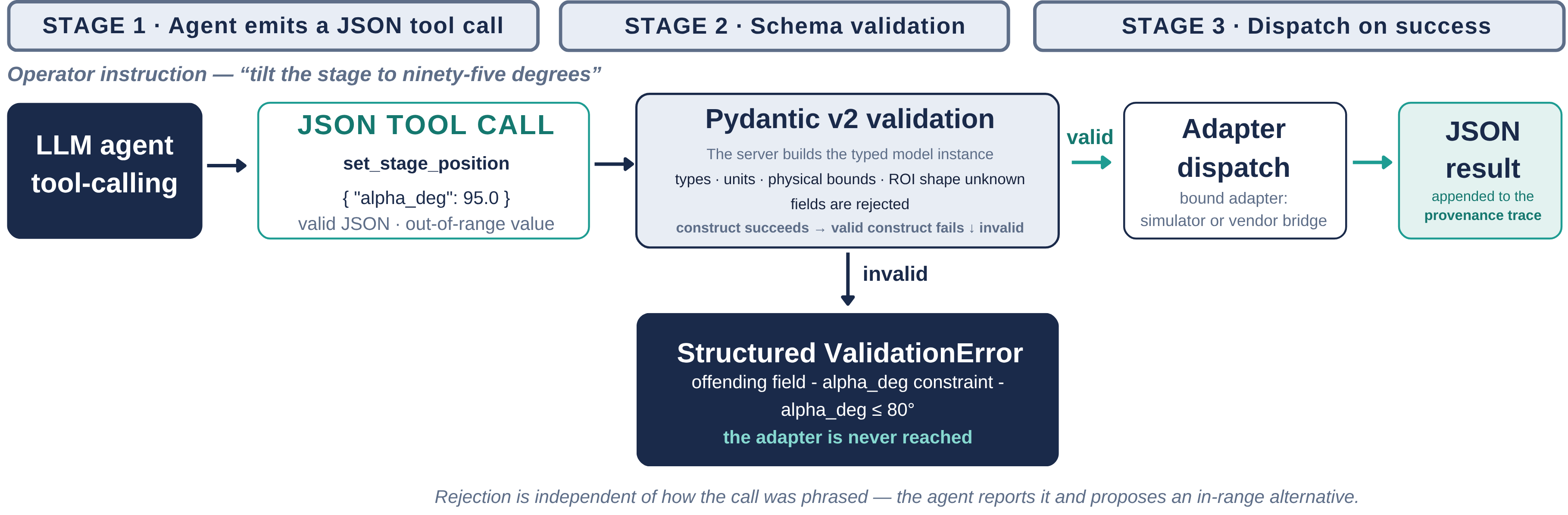}
  \caption{\textbf{Schema-bound validation flow.} A tool invocation proceeds through three discrete stages. The LLM agent emits a JSON
  tool call; the FastMCP server constructs the Pydantic v2 model instance and checks all physical bounds, types, and range constraints. Invalid requests are rejected at the boundary with a structured \texttt{ValidationError} naming the offending field; the adapter is never reached. Valid requests dispatch to the bound
  adapter (simulator or vendor bridge) and return a JSON result that forms part of the provenance trace.}
  \label{fig:schema-flow}
\end{figure*}

Every operation an agent may invoke is exposed as an MCP tool with an associated input schema, expressed as a Pydantic v2 model \citep{Colvin2024Pydantic}. The schema specifies argument names, types, units, default values, and admissible numeric ranges; it forbids unknown fields and rejects malformed compound types (e.g.
regions of interest that are not four-element, monotone, in-image tuples). Validation is performed by the server before any adapter call is
dispatched. A tool invocation therefore proceeds through three discrete stages (Fig.~\ref{fig:schema-flow}); the agent emits a
JSON tool call, the server constructs the Pydantic instance, and only on successful construction is the adapter touched. Failed validation returns a structured \verb|ValidationError| naming the
offending field and constraint; the adapter is never reached. The behaviour is independent of how the LLM phrased the call, so audit traces of rejected requests reflect the agent's actual output
rather than its rationalisation.

We are careful to label this property \emph{bounded execution at the tool boundary}, not safety. The schema enforces physical and operational bounds (e.g.\ $\alpha \in [-80^\circ, +80^\circ]$, dwell-time $\in [0.5,10^4]\,\mu s$, exposure $\in [10^{-3}, 60]$~s). It does not substitute for hardware interlocks, beam-damage models, operator approval, or facility
policy; those layers are discussed under deployment
(Section~\ref{sec:deploy}). A complementary line of work has argued that schema validation is, in itself, a useful basis for trustworthy ML-driven scientific workflows~\citep{RLAM2026}.
\subsection{Vendor-neutral adapter contract}\label{sec:adapter}
\begin{figure*}
  \centering
  \includegraphics[width=\textwidth]{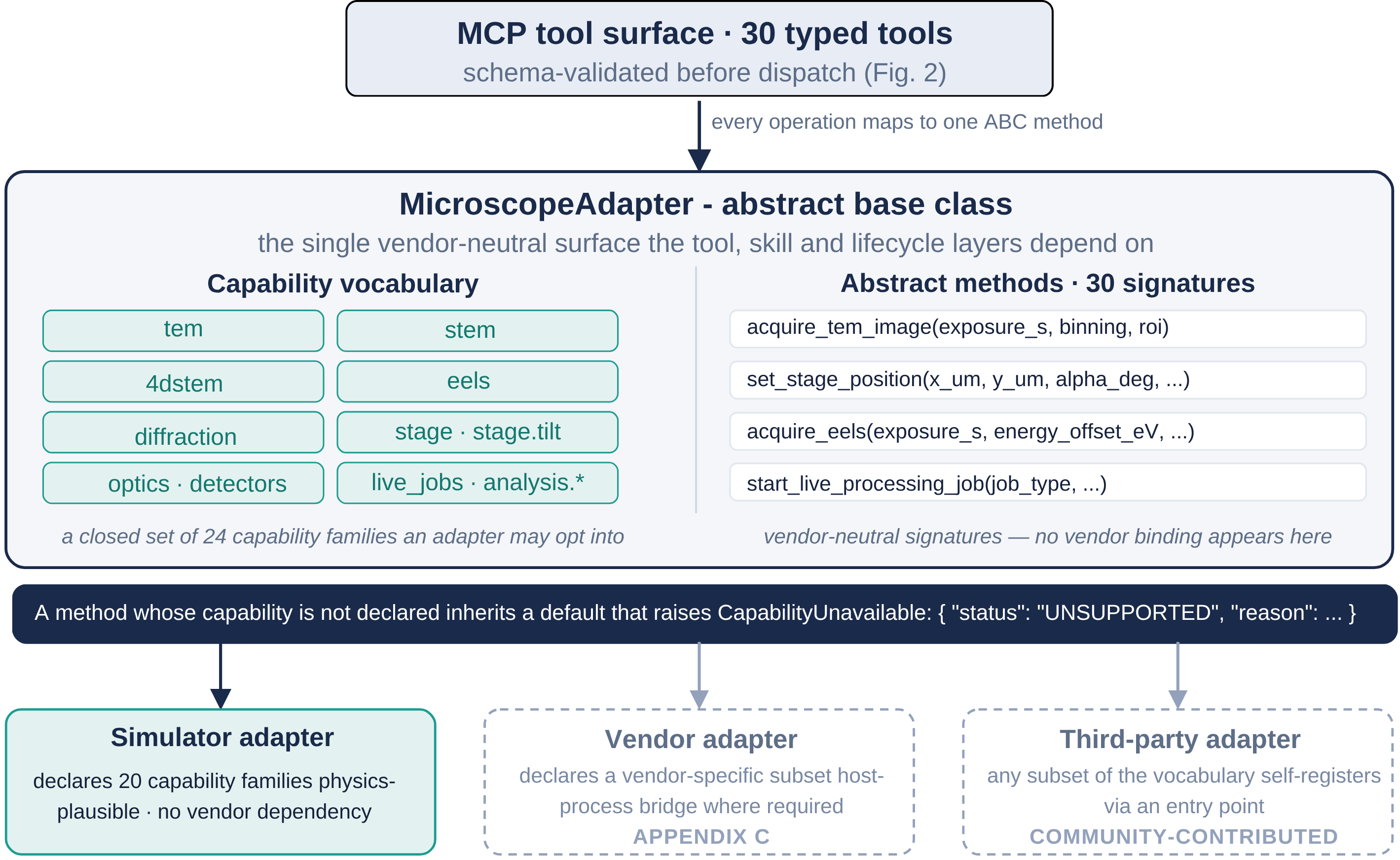}
  \caption{\textbf{Vendor-neutral adapter contract.} The MCP tool
  surface dispatches through the schema layer to the
  \texttt{MicroscopeAdapter} abstract base class. The ABC declares a capability vocabulary and 30 abstract methods with vendor-neutral signatures. Concrete implementations (SimulatorAdapter,
  GatanAdapter, JEOLAdapter, ThermoFisherAdapter, etc..) subclass the ABC and declare only the capabilities they support; undeclared
  capabilities raise \texttt{CapabilityUnavailable} with a structured
  response. Appendix~\ref{app:vendors} gives implementation guidance for adapting the contract to vendor-specific platforms.}
  \label{fig:adapter-contract}
\end{figure*}

The second design choice is the surface across which the schema layer talks to the instrument. We define this surface as a Python abstract base class, \adapter{}, with three properties
(Fig.~\ref{fig:adapter-contract}). First, it declares a \emph{capability vocabulary}: a closed set of 24 capability families that an adapter may opt into. The vocabulary is enumerated in full in Table~\ref{tab:capabilities}; representative families include \cap{tem}, \cap{stem}, \cap{4dstem}, \cap{eels}, \cap{diffraction}, \cap{stage}, \cap{stage.tilt}, \cap{optics}, \cap{detectors}, \cap{live\_jobs}, \cap{analysis.com\_dpc}, and \cap{analysis.radial\_profile}. Second, every operation the server can call is a method of \adapter{} with a vendor-neutral signature. Third, methods whose
capability is not declared inherit a default implementation that raises \texttt{CapabilityUnavailable}; the tool layer translates this into a structured \verb|{"status": "UNSUPPORTED", "reason": ...}| response. An
adapter therefore advertises what it can do (its declared capabilities) and implements only those methods.

This contract has two consequences that matter for the rest of the method. First, the schema, simulator, lifecycle, and skill catalogue
are written against \adapter{} and not against any vendor binding; adding a new instrument is implemented as a subclass and an entry
point. Second, capability negotiation is now part of every session: the agent's first call, by convention, is \tool{get\_capabilities},
which returns the bound adapter's name and capability list. Skills that exercise a particular family begin with a capability check,
which provides a clear failure mode when an adapter does not support a requested family (rather than a silent or misleading error
mid-session). Recent attention to the role of typed tool descriptions in agent reliability supports this design choice~\citep{MCPToolDescriptions2026}. A physics-plausible simulator implements \adapter{} directly, with no vendor dependencies, declaring 20 of the 24 capability families, and is the reference backend exercised in this manuscript. The simulator shares the schema layer with any
live adapter, so the same validation rejects the same arguments regardless of which backend is bound. The hardware-independent test suite runs against the simulator and is bit-stable across machines for a fixed NumPy version.
\subsection{Persistent live-processing jobs}\label{sec:live}
Several useful instrument analyses are not one-shot transformations but long-running, periodically updated computations over a live
front image (e.g. radial-profile tracking,
exponentially-weighted difference imaging, local maximum-FFT mapping, derived filtered views, and 4D-STEM maximum-spot mapping). Exposing each as a single do-the-thing tool would force the agent
to busy-wait or to reissue heavy computations.

\begin{figure*}
  \centering
  \includegraphics[width=\textwidth]{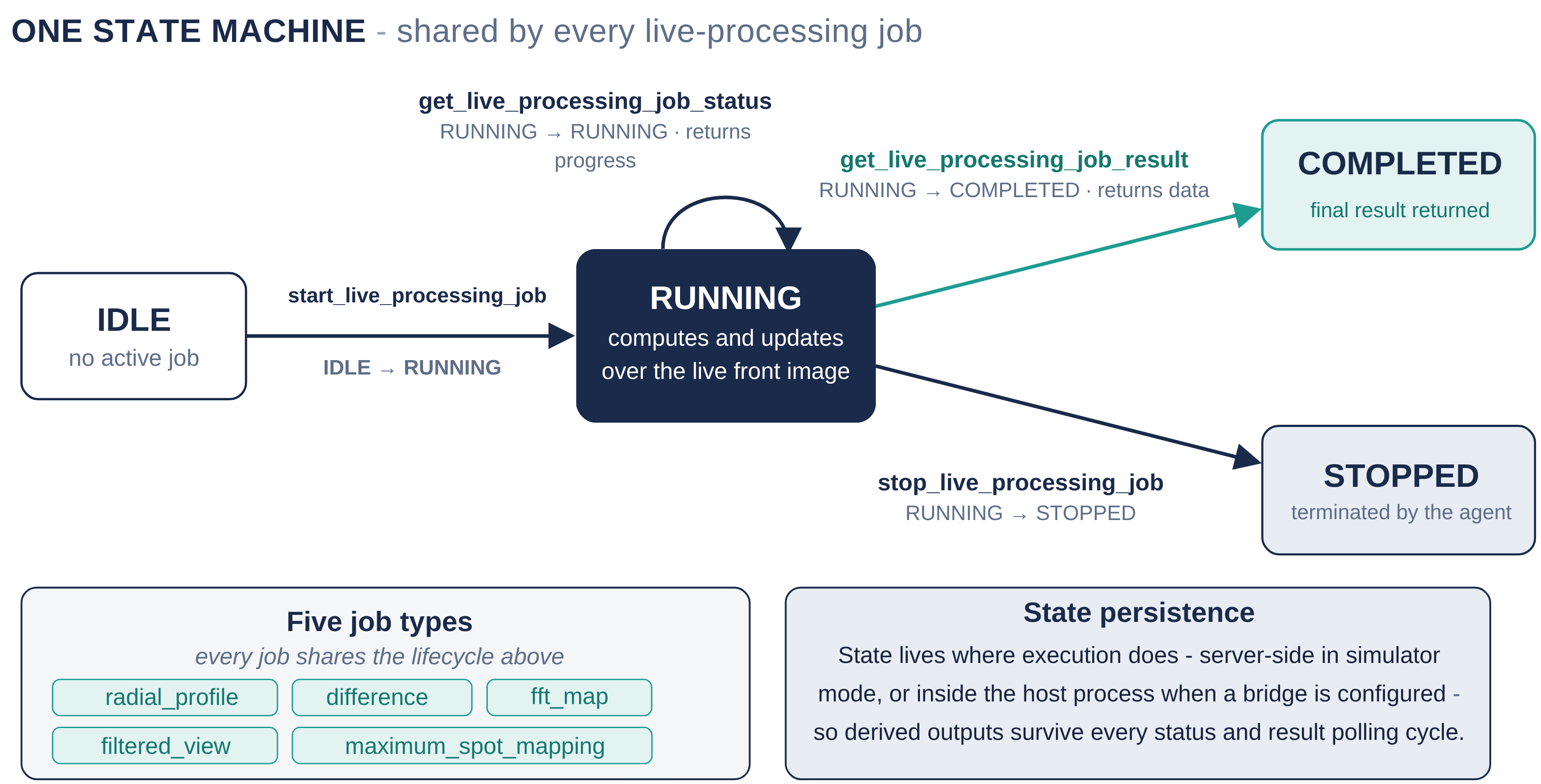}
  \caption{\textbf{Persistent live-processing job lifecycle.} Five live-job types (radial\_profile, difference, fft\_map,
  filtered\_view, maximum\_spot\_mapping) share a common state machine with four transitions: \texttt{start} (IDLE$\to$RUNNING), \texttt{status} (RUNNING$\to$RUNNING, returns progress), \texttt{result} (RUNNING$\to$COMPLETED, returns data), \texttt{stop} (RUNNING$\to$STOPPED). State persists either server-side (simulator mode) or inside the host process (bridge mode), so derived outputs survive across polling cycles.}
  \label{fig:livejob-lifecycle}
\end{figure*}

We promote the lifecycle of a live analysis to a first-class element of the tool surface (Fig.~\ref{fig:livejob-lifecycle}). Four typed tools share a common state machine --- \tool{start\_live\_processing\_job}, \tool{get\_live\_processing\_job\_status}, \tool{get\_live\_processing\_job\_result}, and \tool{stop\_live\_processing\_job} --- and this same lifecycle applies to all five live-job types (\verb|radial_profile|, \verb|difference|, \verb|fft_map|, \verb|filtered_view|, \verb|maximum_spot_mapping|). State lives where execution does: when an implementation-side bridge is configured, jobs are registered inside the host process so derived outputs persist with the rest of the session; otherwise, jobs are held by the server and operate on the adapter's read-only methods.

This factoring has two consequences beyond convenience. First, the agent's record of \texttt{(start, status, result, stop)} calls is itself a machine-readable record of how the analysis evolved, which helps with FAIR-style provenance~\citep{Wilkinson2016FAIR,LLMProvenance2025}. Second, status polling is cheap, so the agent can adapt acquisition parameters in response to live analytical feedback - a closed-loop pattern explored in~\citep{Chen2026EMSeek,Nodeology2025}.
\subsection{Skills: declarative multi-step protocols}\label{sec:skills}
Tools are necessary but not sufficient. A real microscopy session is organised around procedures --- ``run an EELS survey on this specimen'', ``acquire and analyse a 4D-STEM map'', ``align the beam before HRTEM'' --- each of which decomposes into a recurring, partially ordered sequence of tool calls with branching on intermediate results. Asking the LLM to derive these procedures on every session inflates token use, exposes the agent to off-policy mistakes, and produces inconsistent provenance across operators.

\begin{figure*}
  \centering
  \includegraphics[width=\textwidth]{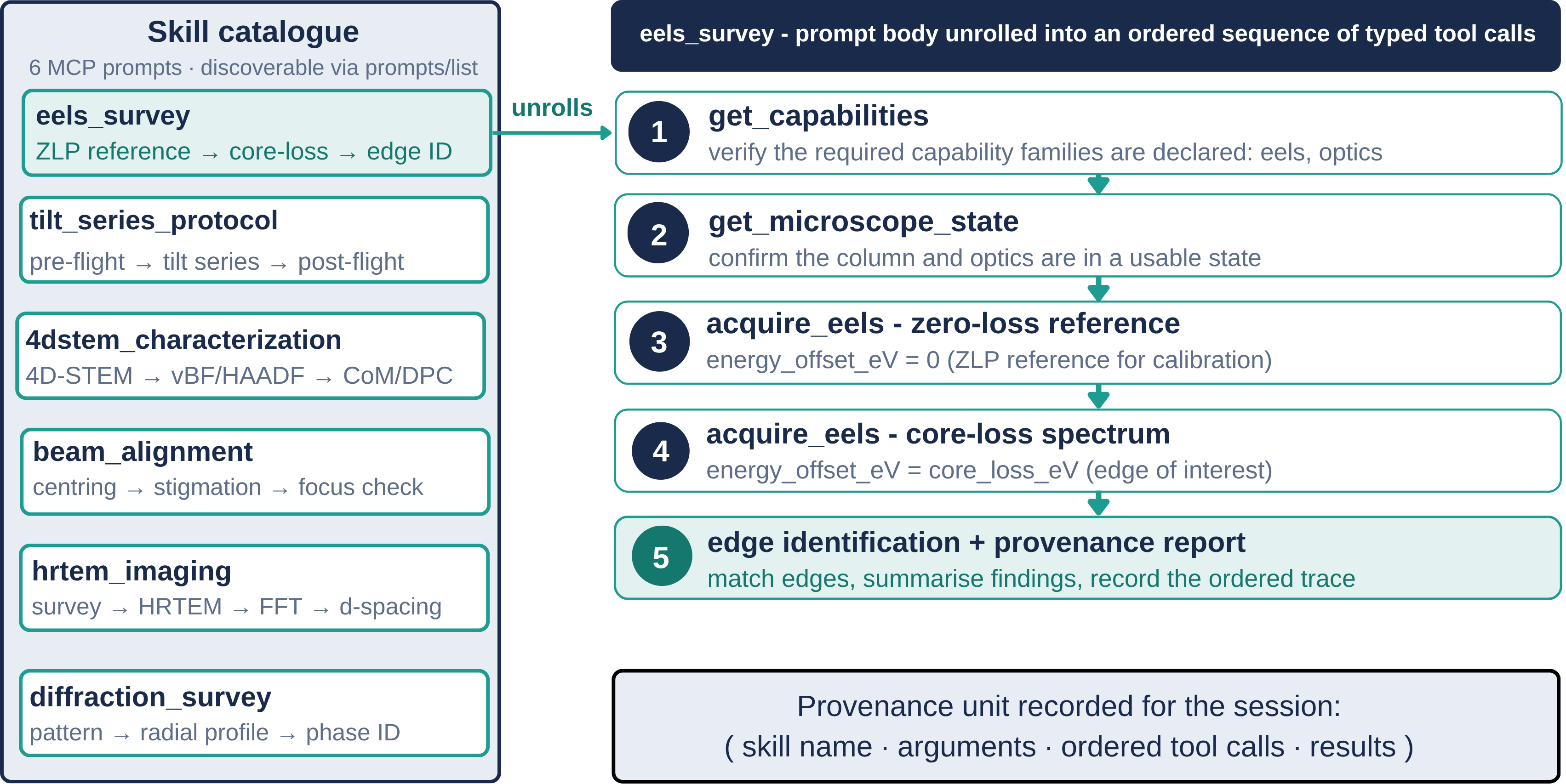}
  \caption{\textbf{Declarative multi-step protocols (``skills'').} Skills are registered as MCP prompts and are therefore discoverable via \texttt{prompts/list}. Each skill composes typed tools from the tool layer into an ordered sequence. The example shows the \texttt{eels\_survey} skill unrolling into: (1)~capability check (\texttt{get\_capabilities}), (2)~microscope-state check (\texttt{get\_microscope\_state}), (3)~ZLP reference acquisition, (4)~core-loss spectrum acquisition, and (5)~edge identification and provenance reporting. All six skills (Table~\ref{tab:skills}) are adapter-portable: the same declarative protocol runs against any adapter that declares the required capabilities.}
  \label{fig:skills-composition}
\end{figure*}

The fourth component of the method addresses this with an abstraction we call a \emph{skill} (Fig.~\ref{fig:skills-composition}). A skill is an implementation pattern, not a new MCP primitive: each skill is a registered MCP \emph{prompt} (one of the three MCP primitives, alongside tools and resources) that returns a structured, parameterised instruction sequence the agent unrolls into ordered tool calls. Each skill carries a name, a short description, and a typed argument list (e.g.\ specimen identity, energy of interest, scan geometry). Because skills sit above the tool layer, they are \emph{adapter-portable}: the EELS-survey skill is the same declarative protocol on every adapter that exposes \cap{eels}. Every skill begins with a \tool{get\_capabilities} check, so an adapter that lacks the required family yields a clear failure rather than an opaque mid-skill error.

Six skills cover the canonical end-to-end workflows we observed across users (Table~\ref{tab:skills}): spectroscopic characterization (\tool{eels\_survey}), tomography (\tool{tilt\_series\_protocol}), 4D-STEM (\tool{4dstem\_characterization}), instrument hygiene (\tool{beam\_alignment}), high-resolution imaging (\tool{hrtem\_imaging}), and crystallographic phase identification (\tool{diffraction\_survey}). Each skill's body encodes the protocol step-by-step: the entry condition, the tool sequence and which fields to populate, the metrics to record, and the report to deliver at the end. The agent's trace is therefore richer than a flat list
of tool calls: it becomes \texttt{(skill, args, tool sequence, results)}, which is the unit of provenance recommended by recent audit-trail work~\citep{LLMProvenance2025,AuditTrails2026}. A
graph-based variant of the same idea has been independently proposed for EM-specific workflows \citep{Nodeology2025}.

\begin{table}[t]
\caption{The six skills registered by the reference implementation. Each skill is a declarative multi-step protocol with typed arguments and a required capability list; the agent unrolls each one into ordered tool calls. The bodies of the skills are vendor-neutral: the same skill runs against any adapter that declares the listed
capabilities.}
\label{tab:skills}
\centering
\small
\setlength{\tabcolsep}{4pt}
\renewcommand{\arraystretch}{1.15}
\begin{tabularx}{\columnwidth}{@{}>{\raggedright\arraybackslash}p{0.30\columnwidth} >{\raggedright\arraybackslash}X >{\raggedright\arraybackslash}p{0.28\columnwidth}@{}}
\toprule
\textbf{Skill} & \textbf{Protocol} & \textbf{Required capabilities} \\
\midrule
\nolinkurl{eels_survey} & ZLP reference; core-loss spectrum; edge identification & \nolinkurl{eels}, \nolinkurl{optics} \\
\nolinkurl{tilt_series_protocol} & Pre-flight; automated tilt series; post-flight check & \nolinkurl{tilt_series}, \nolinkurl{stage} \\
\nolinkurl{4dstem_characterization} & 4D-STEM scan; vBF/HAADF; CoM/DPC; optional spot mapping & \nolinkurl{4dstem}, \nolinkurl{analysis.com_dpc} \\
\nolinkurl{beam_alignment} & Centring; stigmation; focus verification & \nolinkurl{optics}, \nolinkurl{tem} \\
\nolinkurl{hrtem_imaging} & Survey; HRTEM; FFT; $d$-spacing match & \nolinkurl{tem}, \nolinkurl{analysis.radial_profile} \\
\nolinkurl{diffraction_survey} & Diffraction pattern; radial profile; phase ID & \nolinkurl{diffraction}, \nolinkurl{analysis.radial_profile} \\
\bottomrule
\end{tabularx}
\end{table}

\subsection{Modality-specific analyses}\label{sec:analyses}
For 4D-STEM datasets, the tool surface exposes virtual bright- and dark-field maps, a maximum-spot RGB mapping, the centre-of-mass field $\mathbf{CoM}$, and the differential phase contrast magnitude,

\begin{equation}
\mathbf{CoM}(i,j) =
\frac{\sum_{k,l} I(i,j,k,l)\,\mathbf{q}(k,l)}
     {\sum_{k,l} I(i,j,k,l)},
\qquad
\mathrm{DPC}(i,j) = \left\lVert \mathbf{CoM}(i,j) \right\rVert,
\label{eq:com}
\end{equation}

where $I(i,j,k,l)$ is the detector intensity at probe position
$(i,j)$ and detector pixel $(k,l)$, and $\mathbf{q}(k,l)$ is the
corresponding reciprocal-space coordinate, with units of inverse length (typically~\AA$^{-1}$) consistent with the detector calibration. Equation~\eqref{eq:com} therefore returns a reciprocal-space (descriptive) centre-of-mass field; in the simulator this field is physically plausible but is not calibrated quantitative ground truth (Section~\ref{sec:discussion}, Limitations), so it should not be read as a calibrated measurement of the projected electric field. These are the standard
definitions in phase-contrast (S)TEM ~\citep{Ophus20194DSTEM,Savitzky2021py4DSTEM,MullerCaspary2017CoM,Lazic2016iDPC}.
For HRTEM and diffraction patterns, radial profiling and local maximum-FFT mapping are available as separate tools. EELS acquisition takes energy-offset, slit, and dispersion arguments; we delegate quantitative spectral processing to HyperSpy~\citep{Pena2017HyperSpy} and the standard EELS analysis literature~\citep{Egerton2011EELS}.
\subsection{Deployment and bounded-execution model}\label{sec:deploy}
\begin{figure*}
  \centering
  \includegraphics[width=\textwidth]{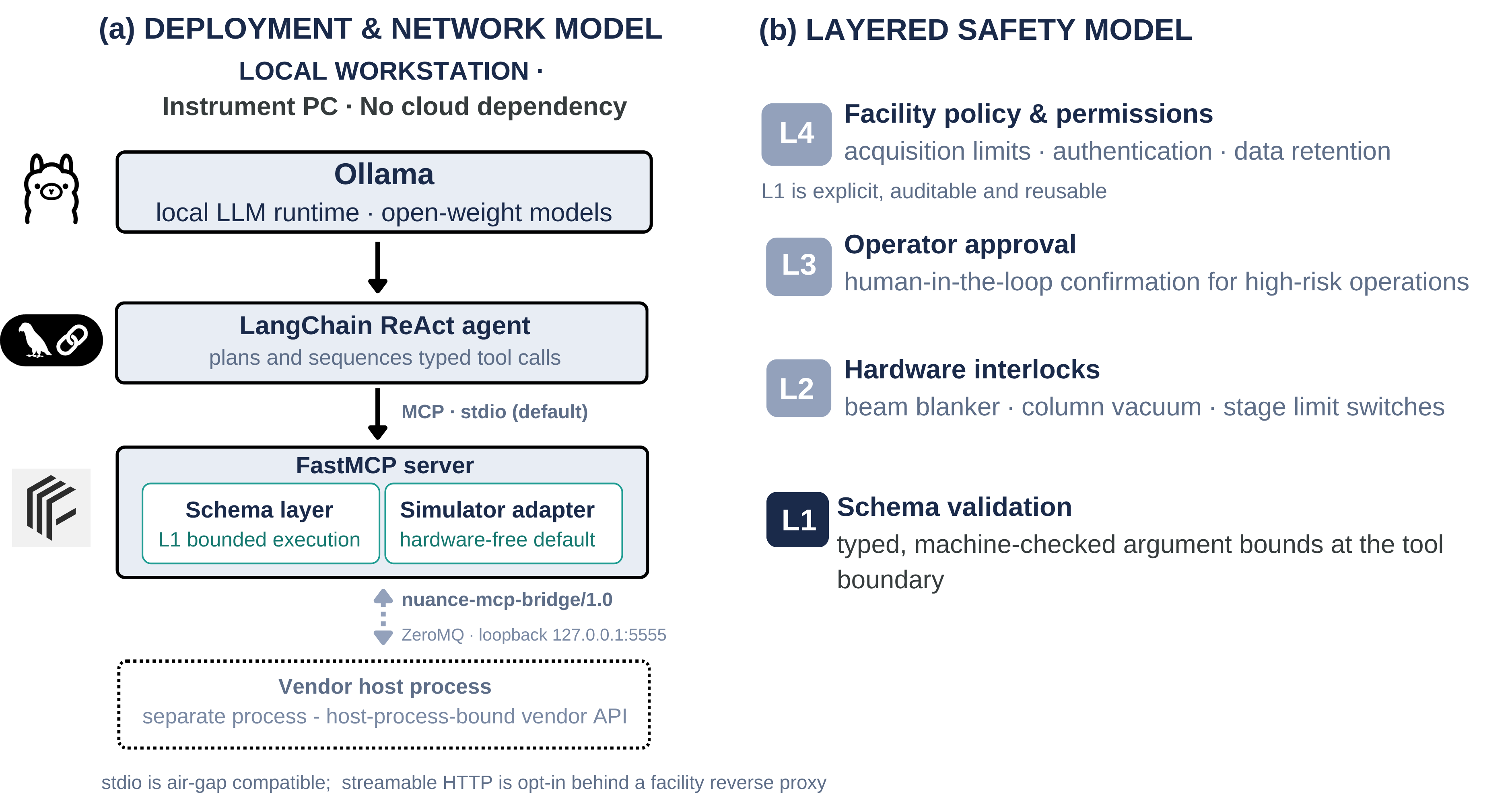}
  \caption{\textbf{Deployment and network model.} The local
  workstation hosts the LLM stack (Ollama), the LangChain ReAct
  agent, and the FastMCP server with its schema layer and simulator.
  When a vendor adapter requires host-process binding, the bridge
  speaks a versioned JSON contract
  (\texttt{nuance-mcp-bridge/1.0}) over ZeroMQ on loopback
  (\texttt{tcp://127.0.0.1:5555}). The layered safety model
  distinguishes four levels: L1~schema validation (this work),
  L2~hardware interlocks, L3~operator approval (human-in-the-loop),
  and L4~facility policy and permissions. Levels are stacked top-to-bottom in the figure (governance at top, schema at bottom) to indicate dependency direction: L1 is the foundation on which the higher layers rest. The default transport is
  stdio; LAN exposure is opt-in.}
  \label{fig:deployment}
\end{figure*}

The framework provides bounded execution at the tool boundary, which is the lowest layer of a layered safety model (Fig.~\ref{fig:deployment}). Schema validation rejects arguments that are out of physical or operational range before any adapter call is dispatched, but it is not equivalent to instrument safety, which requires additional layers: hardware interlocks (beam-blanker
activation, column-vacuum interlocks, stage limit switches, detector-insertion interlocks); operator approval for high-risk
operations through human-in-the-loop confirmation; permission tiers
and authentication; specimen-specific beam-damage and sample-policy
models; and facility-specific gates on acquisition limits, instrument-time windows, and data retention. The contribution of
this work is to make the lowest layer explicit, auditable, and reusable.

The default transport is stdio, suitable for local agent orchestration and fully compatible with air-gapped operation. An optional streamable HTTP transport is provided for cases where a remote MCP-compatible client needs to reach the server through a facility-managed reverse proxy; this transport is not required for the core local-first deployment. Where a vendor API is host-process bound and an adapter therefore needs a bridge plugin, the bridge speaks a versioned JSON contract
(\texttt{nuance-mcp-bridge/1.0}; see Appendix~\ref{app:bridge}) over ZeroMQ \citep{ZMQ2024}, bound by default to \verb|tcp://127.0.0.1:5555| on the instrument PC. Facility-LAN exposure is an opt-in deployment mode protected by host firewall, allow-listed IP ranges, and an HTTPS reverse proxy. The threat model is explicit and limited: named principals are the operator
and the LLM agent process on the workstation; the bridge trusts its caller; loopback is the default; LAN-bound deployments are the
deployer's responsibility to authenticate. In the loopback default, an attacker would require local code execution on the instrument PC to reach the bridge; under opt-in LAN exposure, a misconfigured allow-list or proxy could permit an unauthenticated host on the facility network to issue bridge calls, which is why authentication of LAN-bound deployments is left explicitly to the deployer. Replay protection and rogue-process scenarios are deferred to future work.
\subsection{Validation protocol}\label{sec:validation}
Validation falls into three categories. \textbf{Hardware-independent tests} exercise the schema layer, the simulator-side dispatch, the adapter contract, the live-job lifecycle, and the bridge JSON contract; they require no microscope and are fully deterministic. \textbf{Local-LLM integration tests} drive the same tool surface end-to-end through a real local Ollama \citep{Ollama2024} model and a LangChain ReAct agent \citep{Yao2023ReAct,LangChain2024}, checking that the model selects, parameterises, and sequences tools correctly; outcomes depend on model sampling and are reported separately. \textbf{Bridge structural tests} verify the JSON contract that a vendor-specific adapter plugin must satisfy, and the persistence of live-job state across status/result/stop calls. The v0.2 suite contains 135 tests in total: 120 hardware-independent and 15 local-LLM integration. All 120 hardware-independent tests pass deterministically. The 15 local-LLM tests pass 12--15 of 15 across runs; the three reproducible failures occur under \texttt{qwen2.5:7b} and reflect model non-determinism. The breakdown is given in Table~\ref{tab:tests} and Fig.~\ref{fig:bench}a.

\begin{table}[t]
\caption{Composition of the v0.2 automated validation suite. The
hardware-independent block runs without the microscope, is fully
deterministic, and forms the basis for continuous integration; the
local-LLM block exercises the same tool surface through a real
Ollama model and is reported separately because outcomes depend on
model sampling.}
\label{tab:tests}
\centering
\begin{tabular*}{\columnwidth}{@{\extracolsep\fill}lcc@{\extracolsep\fill}}
\toprule
\textbf{Scope} & \textbf{Tests (pass)} & \textbf{Requirement} \\
\midrule
\multicolumn{3}{l}{\emph{Hardware-independent (deterministic)}} \\[2pt]
Schema validation (Pydantic, ROI/type checks)            & 22 (22)  & none \\
Simulator I/O (modality-specific outputs)                & 32 (32)  & none \\
Tool dispatch and serialisation                          & 38 (38)  & none \\
Live-job lifecycle (start/status/result/stop)            & 15 (15)  & none \\
Bridge protocol (server $\leftrightarrow$ plugin)        & 13 (13)  & none \\
\textbf{Subtotal (deterministic)}                        & \textbf{120 (120)} & Python env \\
\midrule
\multicolumn{3}{l}{\emph{Local-LLM integration (non-deterministic)}} \\[2pt]
End-to-end through Ollama (\texttt{qwen2.5:7b})          & 15 (12--15)\textsuperscript{$\dagger$} & Ollama + model \\
\midrule
\textbf{Total} & \textbf{135 (132--135)} & --- \\
\bottomrule
\multicolumn{3}{p{0.95\columnwidth}}{\footnotesize\textsuperscript{$\dagger$}Pass count varies across runs because model sampling is not deterministic. The three reproducible failures observed in this study are described in Section~\ref{sec:autotests}.} \\
\end{tabular*}
\end{table}
\section{Results}\label{sec:results}
\subsection{Method-induced properties of the protocol}\label{sec:properties}
Three properties of the method are worth isolating because they
derive from design rather than from any one implementation.

\textbf{Adapter-replaceable backend.} Because every adapter implements the same \adapter{} contract (Fig.~\ref{fig:adapter-contract}), the tool layer, the simulator, the live-job lifecycle, and the skill catalogue are unchanged when the adapter is swapped. Concretely, the reference simulator and any future vendor adapter share the same 30 typed tools, six skills (Table~\ref{tab:skills}), and five live-job types (Fig.~\ref{fig:livejob-lifecycle}).

\textbf{Auditable operational bounds.} Every schema rejection is recorded as a structured error tied to the tool name and field (Fig.~\ref{fig:schema-flow}), and the behaviour is independent of how the model phrased the call, consistent with recent reproducibility-constrained large-action-model proposals \citep{RLAM2026}. As a deliberate negative control, a session in which the agent receives the instruction ``tilt the stage to ninety-five degrees'' issues a call to \tool{set\_stage\_position} with \verb|alpha_deg=95.0|. The call is rejected by Pydantic with the message ``\verb|alpha_deg| must be $\leq 80$''; no adapter call is made; the agent reports the rejection to the operator and proposes an in-range alternative.

\textbf{Deterministic regression and CI.} Because the simulator is a stateful, synchronous twin that shares the schema layer with the live path, the same test exercises the same code on every run. The hardware-independent suite therefore runs without microscope time and is reproducible across machines, which addresses the reproducibility gap that LLM-driven scientific software now faces \citep{LLMProvenance2025,RLAM2026}. The same continuous-integration pipeline runs a project-status probe that re-measures the protocol surface --- typed tools, skills, and live-job types --- directly from the running server via \verb|tools/list| and \verb|prompts/list| on every push, so the surface counts reported in this manuscript cannot drift from the code without failing CI (Appendix~\ref{app:evolution}). The complete 30-tool surface is enumerated in Table~\ref{tab:tools} (Appendix~\ref{app:evolution}).
\subsection{Automated validation outcomes}\label{sec:autotests}
In v0.2, all 120 hardware-independent tests pass deterministically (Table~\ref{tab:tests}, Fig.~\ref{fig:bench}a). The block executes in approximately 18~s on commodity workstations and forms the basis of continuous integration. The 15 local-LLM integration tests pass between 12 and 15 of 15 across runs; under \texttt{qwen2.5:7b} on the testbed used here, three of the 15 tests fail reproducibly because the model occasionally re-orders steps in a multi-call sequence. The failures are recorded as part of the validation budget rather than suppressed, since they characterise the LLM rather than the protocol. The integration suite confirms that an off-the-shelf local model can select, parameterise, and sequence tools through the schema-bound surface in realistic microscopy task sequences (e.g.\ ``acquire a HAADF image, start a radial-profile live job, report when the dominant ring shifts''). All operations in these sequences dispatch to the reference simulator rather than to a live instrument.
\subsection{Same-hardware local-LLM probe}\label{sec:probe}
\begin{figure*}
\centering
\includegraphics[width=\textwidth]{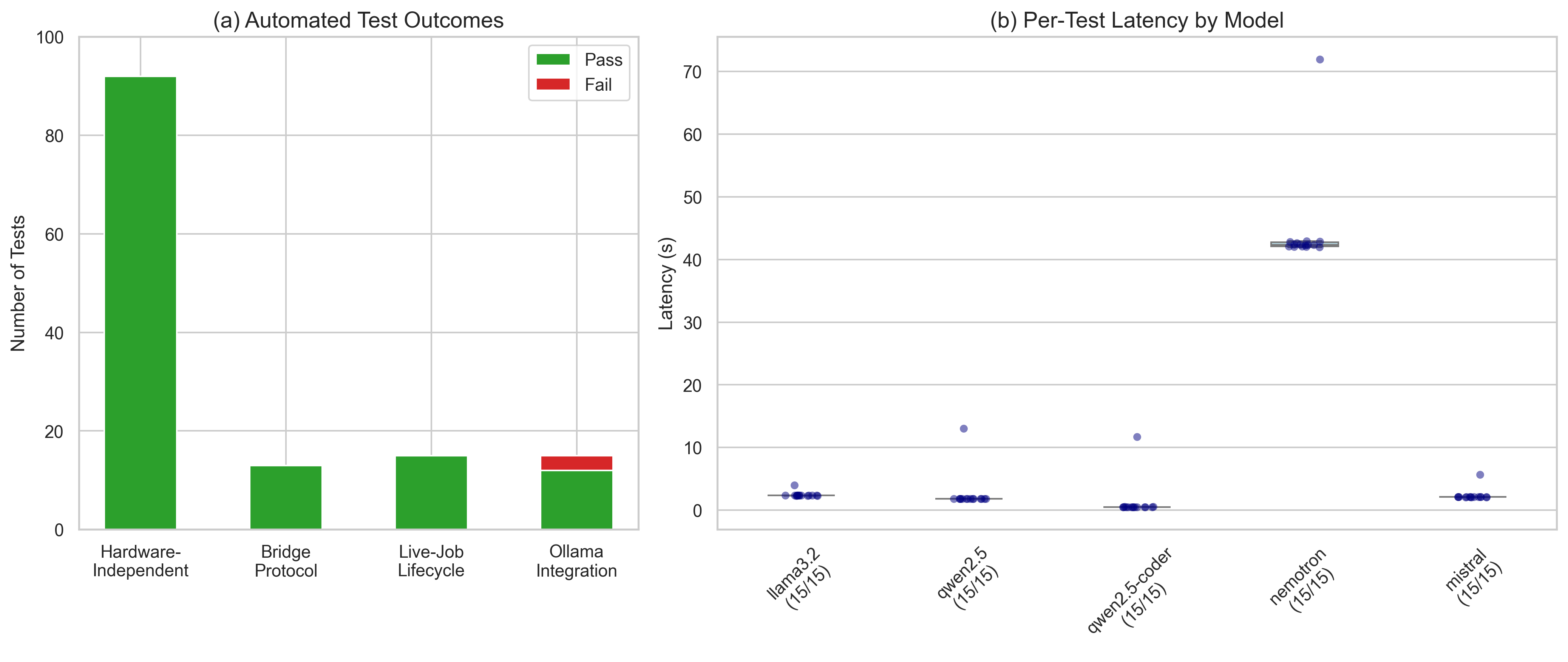}
\caption{\textbf{Validation behaviour of the v0.2 release against the reference simulator.}
\textbf{(a)}~Pass/fail counts of the v0.2 automated suite by category. The hardware-independent, bridge-protocol, and live-job lifecycle blocks are
fully deterministic and pass in their entirety. The local-LLM (Ollama) block runs under \texttt{qwen2.5:7b} and fails three of fifteen tests reproducibly; the failures reflect model
non-determinism in multi-call sequences rather than schema or adapter behaviour.
\textbf{(b)}~Per-test latency distributions on a same-hardware Apple Silicon testbed (May~2026) for the 15-test local-LLM suite across five openly available tool-calling models, all of which
pass 15/15 on this single-run probe. Median latencies sit between 1~s and 3~s for \texttt{llama3.2}, \texttt{qwen2.5},
\texttt{qwen2.5-coder}, and \texttt{mistral}, and near 42~s for \texttt{nemotron}. The probe is not replicated and confidence intervals are not reported; we caution against quantitative
inter-model comparison from this dataset alone
(Section~\ref{sec:probe}).}
\label{fig:bench}
\end{figure*}

We do not attempt cross-platform LLM benchmarking, which is not meaningful across heterogeneous accelerators and on which the Berkeley Function-Calling Leaderboard provides a rigorous treatment
\citep{Patil2025BFCL}. As a single-run probe on a same-hardware Apple Silicon testbed (May~2026), the 15-test local-LLM integration suite was executed against five openly available models with
tool-calling support in Ollama: \texttt{llama3.2}, \texttt{qwen2.5}, \texttt{qwen2.5-coder}, \texttt{nemotron}, and \texttt{mistral}. All
five passed 15/15 on this run (Fig.~\ref{fig:bench}b). For \texttt{qwen2.5} this is at the upper end of the 12--15/15 range
reported in Section~\ref{sec:autotests}, consistent with the model's non-determinism across runs. Median per-test latencies separated into two regimes: roughly 1-3~s for four of the five models, and approximately 42~s for \texttt{nemotron}, with one outlier near 75~s. Models without tool-calling support in their
current Ollama builds were excluded.

The qualitative point for the method is that the schema-bound tool surface is sharp enough that small open-weight models from four different providers (Meta, Alibaba, NVIDIA, and Mistral AI) can drive it on local hardware, without cloud dependency. The inter-model latency spread is wide enough that model choice matters in practice even when correctness does not. This single run is undersampled, and we caution against reading too much into it; replicated benchmarking with confidence intervals and BFCL-style methodology is future work.

\section{Discussion}\label{sec:discussion}
The method's principal contribution is reproducible control semantics for agentic instrumentation at the level of the protocol itself. Schema-bound tool typing makes the boundary between agent intent and instrument action explicit and machine-checkable (Fig.~\ref{fig:schema-flow}); the vendor-neutral adapter contract keeps that boundary thin and replaceable (Fig.~\ref{fig:adapter-contract}); the persistent live-job lifecycle exposes long-running, stateful analyses through the same typed interface as single-shot operations (Fig.~\ref{fig:livejob-lifecycle}); and the skill abstraction lifts recurring multi-tool procedures into MCP-prompt-registered artefacts that are discoverable through ordinary protocol calls (Fig.~\ref{fig:skills-composition}). Together these four choices give a runtime in which agent traces are analyzable, regressions are deterministic, and the entire layer above the adapter ports across vendors. Recent ``thinking microscope'' framings cast the instrument itself as an iterative reasoner over its own data \citep{Jamali2026ThinkingMicroscopes}; the contribution here is the protocol surface on which that view can be made operational and auditable.

The shape of the contribution differs from a hardware demonstration. We do not claim improved imaging or new physics. We claim that once the control surface is right, agentic workflows demonstrated piecemeal in chemistry \citep{Boiko2023Coscientist,Bran2024ChemCrow,Tom2024SDL}, scanning probes \citep{Kalinin2021AutoEM,Liu2025RewardSPM,Mandal2025AILA}, and TEM \citep{Wall2025TEMAgent,Chen2026EMSeek,Jiang2025Ptycho,Nodeology2025} can be composed out of typed tools with bounded execution, declarative skills, and a reproducible runtime.

The 2026 Nature reports on Co-Scientist, Robin, and Empirical Research Assistance sharpen this point rather than weakening it \citep{Gottweis2026CoScientist,Ghareeb2026Robin,Aygun2026ERA}. They show that multi-agent systems can already operate over scientific knowledge, code, candidate hypotheses, and human-executed experiments. Microscopy adds a harder control layer: the next action must be executable on the bound instrument, safe for the loaded sample, meaningful under the current calibration state, and interpretable under a measurement model. In that setting a language-model agent needs a calibrated digital twin of the instrument and sample workflow rather than a chat interface to a vendor API. The simulator and schema-bound MCP surface presented here move toward that digital-twin layer by turning instrument actions into typed, testable, replayable objects that an agent can reason about before a physical experiment is attempted.

The simulator has independent value. Because it shares the schema layer with any live adapter, hardware-independent tests cover the same code that would run against an instrument, so workflow logic can be developed and audited without instrument time. This is a recurring obstacle in microscope-software development \citep{Mastronarde2005SerialEM}; sharing the schema with the simulator removes it for the validation suite.

The deployment model (Fig.~\ref{fig:deployment}) makes the security
boundary explicit. The bridge trusts its caller on loopback; LAN
exposure is opt-in and the deployer's responsibility. This is a
deliberate scope limit: the protocol provides the typed, auditable
lowest layer of a layered safety model, but does not substitute for
hardware interlocks, operator approval, or facility policy. The
vendor adapter comparison in Table~\ref{tab:vendor-comparison}
(Appendix~\ref{app:vendors}) shows that the bridge pattern
generalises across four major vendor platforms with different
scripting surfaces and host-binding mechanisms.
\subsection*{Positioning relative to concurrent work}
Table~\ref{tab:sota} situates the method against the closest concurrent systems. The dimensions tabulated are the design goals of this work plus two further axes; the comparison is therefore one in which competing systems are scored against features this work set out to provide, and a ``\xmark'' marks a feature not reported in the cited source rather than a deficiency of the system on its own terms. To keep the comparison even-handed, we include a live-instrument-validation row, an axis on which this work does \emph{not} lead: AILA \citep{Mandal2025AILA} and \citet{Chen2026EMSeek} report demonstrations on physical instruments, whereas \nmcp{} is validated only in software here (Section~\ref{sec:results}; Limitations). \temagent{} \citep{Wall2025TEMAgent} demonstrated that MCP-mediated TEM workflows are feasible but exposes a vendor-specific tool surface without schema-bound validation or an adapter abstraction. stemOrchestrator \citep{Pratiush2025stemOrchestrator} provides a hardware-abstraction layer for STEM detectors and cameras but operates below the MCP protocol level and does not expose typed tools or declarative skills to an LLM agent. AILA \citep{Mandal2025AILA} automates atomic-force microscopy through LLM agents with real-instrument demonstrations but uses a bespoke tool interface rather than MCP. The broader MCP-for-science survey of \citet{Pan2025MCPScience} and the MCP-Bench evaluation framework \citep{Wang2025MCPBench} confirm growing interest in standardised tool protocols for scientific workflows, while \citet{Liang2026MCPChemEng} have extended MCP to chemical process simulation.

The combination of schema-bound validation, a vendor-neutral adapter contract, a persistent live-job lifecycle, and a prompt-registered skill abstraction is, to our knowledge, not jointly present in the systems we surveyed. The principal gap relative to production-grade platforms is security: commercial offerings provide OAuth~2.0, 21~CFR~Part~11 compliance, and fine-grained audit trails, whereas the bridge security model presented here is intentionally minimal (Section~\ref{sec:deploy}).

\begin{table}[t]
\caption{Comparison of the method with concurrent MCP and
agentic-instrumentation systems, scored against publicly reported
features. The tabulated dimensions are the four design goals of
this work plus two further axes (a physics-plausible simulator and
an authenticated security model), so the comparison is framed
around features this work set out to provide; a live-instrument
validation (LV) row is included as a dimension on which this work
does not lead. Columns indicate whether each system, as documented
in its public sources, provides schema-bound validation (SV), a
vendor-neutral adapter contract (AC), a live-job lifecycle (LJ), a
declarative skill abstraction (SK), a physics-plausible
simulator (Sim), an authenticated security model beyond
loopback trust (Sec), and a reported live-instrument validation
(LV). ``\xmark'' denotes the feature is not
reported in the cited source; ``n/a'' denotes that the dimension
does not apply to that class of system; ``min.'' denotes the
intentionally minimal model of this work (loopback by default;
LAN authentication is the deployer's responsibility;
Section~\ref{sec:deploy}). Feature taxonomies of comparable systems
may evolve; this table reflects information available at
submission.}
\label{tab:sota}
\centering
\small
\begin{tabular*}{\columnwidth}{@{\extracolsep\fill}lccccccc@{\extracolsep\fill}}
\toprule
\textbf{System} & \textbf{SV} & \textbf{AC} & \textbf{LJ} & \textbf{SK} & \textbf{Sim} & \textbf{Sec} & \textbf{LV} \\
\midrule
\nmcp{} (this work)                                   & \cmark & \cmark & \cmark & \cmark & \cmark & min.   & \xmark \\
\temagent{} \citep{Wall2025TEMAgent}                  & \xmark & \xmark & \xmark & \xmark & \xmark & \xmark & \cmark \\
stemOrchestrator \citep{Pratiush2025stemOrchestrator} & \xmark & \cmark & \xmark & \xmark & \xmark & \xmark & \cmark \\
AILA \citep{Mandal2025AILA}                           & \xmark & \xmark & \xmark & \xmark & \xmark & \xmark & \cmark \\
MCP-for-Science \citep{Pan2025MCPScience}             & \xmark & \xmark & \xmark & \xmark & \xmark & \cmark & n/a    \\
MCP-Bench \citep{Wang2025MCPBench}                    & n/a    & n/a    & n/a    & n/a    & n/a    & n/a    & n/a    \\
\bottomrule
\end{tabular*}
\end{table}
\subsection*{Limitations}
The validation in this manuscript is software-only: every claim is demonstrated against the physics-plausible simulator that implements the same \adapter{} contract. Live-instrument demonstrations on commercial scientific hardware are intentionally out of scope here; they are described separately under the corresponding intellectual-property disclosure \citep{dosReis2025Disclosure}. Because that live-instrument validation is withheld under the disclosure, on-hardware performance of the method cannot be independently assessed from this manuscript; the software-only scope is therefore both a deliberate design choice and a constraint on what readers can conclude. The local-LLM probe in Section~\ref{sec:probe} is a single-run result across five models; reported pass rates and latencies are not accompanied by confidence intervals and should be treated as indicative rather than quantitative. The bridge security model is intentionally minimal: loopback by default, with LAN exposure left to the deployer to authenticate; replay protection and rogue-process scenarios are not currently addressed. The simulator outputs are physically plausible but are not calibrated quantitative ground truth. Each of these scope limits maps onto a near-term work item rather than a challenge to the method itself.
\section{Conclusion}\label{sec:conclusions}
We have described a vendor-agnostic method for connecting local LLM agents to scientific instruments at the level of the tool protocol, and presented its reference implementation as a software-only validation study. The method combines a schema-bound tool surface (Fig.~\ref{fig:schema-flow}), a vendor-neutral adapter contract (Fig.~\ref{fig:adapter-contract}) paired with a physics-plausible simulator, a persistent live-processing job lifecycle (Fig.~\ref{fig:livejob-lifecycle}), and a skill abstraction implemented through MCP prompts (Fig.~\ref{fig:skills-composition}, Table~\ref{tab:skills}) that composes typed tools into reusable multi-step protocols. The reference implementation \nmcp{} exposes 30 typed tools, 5 live-job types, and 6 skills. All validation reported here is software-only, against the reference simulator: all 120 hardware-independent tests pass deterministically, and the 15 local-LLM integration tests pass 12--15 of 15 across runs, with the three reproducible failures attributable to model non-determinism (Table~\ref{tab:tests}). The deployment model (Fig.~\ref{fig:deployment}) makes the security boundary explicit and the layered safety model auditable. The MIT-licensed source, simulator, and test suite are available at \url{https://github.com/NUANCE-IT/nuance-mcp}.

The protocol is intended as a working foundation for closed-loop and autonomous microscopy research: schema-checked agent traces, reproducible regression suites, and adapter-portable skills can be composed against it without re-solving the host-process binding, governance, and bounded-execution frictions identified in Section~\ref{sec:intro}.

\section*{Competing interests}
The authors declare the following competing interests. Certain aspects of the broader universal MCP framework for scientific instrumentation described in this work are the subject of a pending U.S. patent application filed on 13 July 2026. Related technology was previously disclosed to Northwestern University as Invention Disclosure Disc-ID-25-05-22-002 (technology ID 2025-136). The inventors are R. dos Reis and V. P. Dravid, and Northwestern University is the assignee. The authors may financially benefit from patents or licensing agreements arising from this intellectual property.
\section*{Author contributions}
R.d.R.\ conceived the architecture, designed and implemented the
schema-bound tool protocol, the adapter contract, the bridge
specification, the simulator, the live-job lifecycle, and the skill
catalogue, implemented the full software stack and validation
suite, and drafted the manuscript. V.P.D.\ supervised the project,
contributed scientific framing, and revised the manuscript. All authors read and approved the final manuscript.
\section*{Acknowledgments}
This work made use of the NU\textit{ANCE} facility (RRID: SCR\_026361) at Northwestern University, which has received support from the IIN and Northwestern's MRSEC program (NSF DMR-2308691). The authors acknowledge the open-source communities behind FastMCP, MCP, Ollama, LangGraph,
\texttt{faster-whisper}, and the broader scientific-Python ecosystem. The authors thank Benjamin Miller, Anahita Pakzad, and Stephen Mick from Gatan/AMETEK for their early testing and for discussions regarding the implementation of NUANCE-MCP within the GMS environment.
\section*{Data availability}
No experimental dataset is required to reproduce the software
validation reported here. All results in this manuscript are
reproducible against the physics-plausible simulator that ships
with the source distribution and the accompanying test suite.

\section*{Code availability}
Source code, tests, examples, and documentation are available
under the MIT License at:
\url{https://github.com/NUANCE-IT/nuance-mcp}.
\bibliographystyle{unsrtnat}
\bibliography{reference}

\newpage
\section*{Appendix}
\appendix
\section{Bridge protocol \texttt{nuance-mcp-bridge/1.0}}\label{app:bridge}
This appendix specifies the versioned JSON contract that
host-process-bound adapter implementations must satisfy. The
contract is vendor-neutral; specific adapter implementations and
their results are intentionally out of scope for this manuscript.
The bridge speaks a versioned JSON contract over a ZeroMQ
\verb|REQ|/\verb|REP| pair. Each request carries a version envelope
(\verb|v: "nuance-mcp-bridge/1.0"|), a method name matching an
\adapter{} method, and a parameter object. The response carries a
status code (\verb|ok|, \verb|error|, \verb|unsupported|), a result
payload, and a structured error string when appropriate. The first
call after connect must be \verb|hello|, which negotiates the
protocol version and returns the bridge's capability list; clients
reject unknown version prefixes. The bridge pumps the vendor's GUI
event loop at least every 100~ms while waiting, so the host
application remains responsive during long polls.
Image payloads use a \verb|name|/\verb|shape|/\verb|data_dtype|/
\verb|data_b64| encoding; spectra use a parallel
\verb|counts_b64|/\verb|energy_eV| pair. New optional fields can be
added at any time; receivers must ignore unknown fields. Breaking
changes increment the version prefix and trigger an explicit
handshake failure with older partners. The full schema and the
30-tool surface that adapters expose are documented in the
\texttt{docs/spec/} directory of the source repository.
\section{Evolution from v0.1.0 to v0.2}\label{app:evolution}
The protocol-level method described in this manuscript did not arrive fully formed. Three release lines preceded it. We summarise the trajectory here because each step illustrates a design decision that the current method makes explicit.
\paragraph{v0.1.0: monolithic server, ad-hoc validation.}
The first release shipped as a single FastMCP server with 21 tools
hand-written against one host-process API. Argument validation
existed but was scattered through individual tool bodies; out-of-range
values often produced opaque vendor-side errors rather than
structured rejections. There was no abstract base class: the tool
layer talked to the instrument through a direct import. The
simulator (a separate file with hand-written modality kernels)
existed but was selected by an environment flag rather than as a
peer adapter, and it did not share the validation path with the
live code. There were no skills; multi-step procedures lived in
example scripts that operators copied and adjusted by hand.
\paragraph{v0.1.1: centralized schema, host-process bridge.}
Two changes consolidated the v0.1.0 ad-hocery. First, every tool
gained a Pydantic v2 input model and the dispatcher validated those
models before any vendor call, which made rejection paths
structured and prompt-independent for the first time. Second, the host-process binding was crossed by a small JSON-over-ZeroMQ ``bridge'' plugin that ran inside the vendor process; the FastMCP server became an ordinary Python process again. The bridge was unversioned: client and server agreed on the message shape by implicit convention, which became a maintenance problem as the schema grew.
\paragraph{v0.1.2: live-processing lifecycle, expanded tool set.} This release introduced the persistent live-processing job abstraction - start, poll, fetch result, stop - as four typed tools sharing a common state machine, with five live-job types (\verb|radial_profile|, \verb|difference|, \verb|fft_map|,
\verb|filtered_view|, \verb|maximum_spot_mapping|). The tool count grew from 21 to 30 with the addition of fine-grained optics, beam, and detector controls that earlier releases had bundled into coarser surfaces. The automated test suite grew in step from 78 to 107 deterministic tests; the local-LLM integration block (15 additional tests exercising the same tool surface through Ollama) is added later in v0.2 (Section~\ref{sec:autotests}). The simulator still lived alongside the host-process binding rather than as a peer. 
\paragraph{v0.2.0: protocol-level method, adapter contract,
versioned bridge, skills.} The current release reorganises the preceding pieces around four protocol-level abstractions described in Section~\ref{sec:methods}. The schema layer becomes the single validation surface (no per-tool validation logic remains). The \adapter{} abstract base class is introduced as the only surface the schema layer talks to; the simulator now implements that contract as a peer adapter rather than as a fallback path. The bridge protocol is versioned (\texttt{nuance-mcp-bridge/1.0}) and includes an explicit \verb|hello| handshake that negotiates capability vocabulary. Six declarative \emph{skills} are added as MCP prompts that compose typed tools into reusable multi-step protocols (Table~\ref{tab:skills}). Default network exposure is tightened from all-interfaces to loopback only, and a layered safety model is documented so that the schema layer is no longer conflated with instrument safety (Fig.~\ref{fig:deployment}).
Table~\ref{tab:evolution} summarises the trajectory.
\paragraph{Development tooling - reproducible project status.} Independently of the protocol surface, the v0.2 line adds a single-source-of-truth project-status probe
(\texttt{scripts/project\_status.py}). The probe measures the live protocol surface - typed tools, skills, live-job types, and capabilities - directly from the running server through \texttt{tools/list} and \texttt{prompts/list} rather than from documentation, runs the test suite, reads version-control state, and emits a milestone roadmap. A single implementation is surfaced three ways: as a command-line report, as a \texttt{Project status probe} step in continuous integration, and as a \texttt{/goals} slash command for editor-integrated agents. It is deliberately \emph{not} part of the instrument-control tool surface - it adds nothing to the 30 typed tools - but it is a direct consequence of the schema-bound framing: because the tool surface is introspectable, project status becomes a measurement rather than a claim, and the surface figures quoted throughout this manuscript are re-checked
against the code on every continuous-integration run.

\begin{table}[h]
\caption{Cumulative evolution of the protocol surface from v0.1.0
to v0.2. Each row marks the release in which the corresponding
abstraction first appeared; later releases inherit and extend
prior ones.}
\label{tab:evolution}
\centering
\small
\begin{tabularx}{\columnwidth}{@{}>{\raggedright\arraybackslash}X c c c >{\raggedright\arraybackslash}X@{}}
\toprule
\textbf{Component} &
  \textbf{v0.1.0} & \textbf{v0.1.1} & \textbf{v0.1.2} & \textbf{v0.2} \\
\midrule
Typed tools (count)                        & 21          & 21          & 30           & 30 \\[2pt]
Per-tool schema validation                 & ad hoc      & centralized & centralized  & centralized \\[2pt]
Host-process bridge                        & ---         & unversioned & unversioned  & versioned 1.0 \\[2pt]
Live-processing job lifecycle              & ---         & ---         & 5 types      & 5 types \\[2pt]
Vendor-neutral adapter ABC (\adapter{})    & ---         & ---         & ---          & \cmark \\[2pt]
Capability vocabulary                      & ---         & ---         & ---          & \cmark \\[2pt]
Skills (MCP prompts)                       & ---         & ---         & ---          & 6 skills \\[2pt]
Default network exposure                   & all ifaces  & all ifaces  & all ifaces   & loopback \\[2pt]
Automated test count (passing/total)       & 65/68       & 78/82       & 107/107      & 120/120 (det.) + 12--15/15 (LLM) \\[2pt]
Layered safety model documented            & ---         & ---         & ---          & \cmark \\[2pt]
Project-status probe (\texttt{/goals}, CI) & ---         & ---         & ---          & \cmark \\
\bottomrule
\end{tabularx}
\end{table}
Two threads run through the table that are worth calling out explicitly. First, every later abstraction was prefigured as ad-hoc code in an earlier release: validation existed before it was centralized, bridge messaging existed before it was versioned, the simulator existed before it was made a peer adapter. The protocol work was about \emph{naming} these surfaces, not about inventing them from scratch. Second, the test count grows monotonically with the surface, and the v0.2 line is the first to expose model non-determinism by including a local-LLM integration block. Reporting those failures as part of the validation budget --- rather than as flakiness to be hidden --- is a deliberate consequence of the schema-bound framing.

For completeness, and to let a reader verify the surface counts quoted throughout the manuscript against the code, Table~\ref{tab:capabilities} enumerates the full 24-family capability vocabulary and Table~\ref{tab:tools} enumerates the 30 typed tools. Both tables are generated from the same \texttt{tools/list} and \texttt{prompts/list} introspection used by the continuous-integration probe.

\begin{table}[h]
\caption{The 24-family capability vocabulary declared by the
\adapter{} abstract base class. An adapter opts into a subset; the
reference simulator declares 20 of the 24 families (those not
marked ``vendor-only''). Families group into instrument modalities,
hardware subsystems, live-processing, and analysis routines.}
\label{tab:capabilities}
\centering
\small
\setlength{\tabcolsep}{4pt}
\renewcommand{\arraystretch}{1.15}
\begin{tabularx}{\columnwidth}{@{}>{\raggedright\arraybackslash}p{0.26\columnwidth} >{\raggedright\arraybackslash}X@{}}
\toprule
\textbf{Group} & \textbf{Capability families} \\
\midrule
Imaging modalities & \nolinkurl{tem}, \nolinkurl{stem}, \nolinkurl{4dstem}, \nolinkurl{eels}, \nolinkurl{diffraction}, \nolinkurl{edx} \\
Stage \& geometry  & \nolinkurl{stage}, \nolinkurl{stage.tilt}, \nolinkurl{stage.rotation}, \nolinkurl{tilt_series} \\
Optics \& beam     & \nolinkurl{optics}, \nolinkurl{optics.aberration}, \nolinkurl{beam}, \nolinkurl{beam.blanker} \\
Detectors          & \nolinkurl{detectors}, \nolinkurl{detectors.haadf}, \nolinkurl{detectors.camera} \\
Live processing    & \nolinkurl{live_jobs} \\
Analysis           & \nolinkurl{analysis.com_dpc}, \nolinkurl{analysis.radial_profile}, \nolinkurl{analysis.fft}, \nolinkurl{analysis.virtual_aperture}, \nolinkurl{analysis.peak_find}, \nolinkurl{analysis.dspacing} \\
\bottomrule
\end{tabularx}
\end{table}

\begin{table}[h]
\caption{The 30 typed tools exposed by the reference
implementation, grouped by function. Each tool maps to one
\adapter{} method with a vendor-neutral signature and is validated
by its Pydantic v2 input model before dispatch
(Section~\ref{sec:schema}). Counts are re-measured from the running
server on every continuous-integration push.}
\label{tab:tools}
\centering
\small
\setlength{\tabcolsep}{4pt}
\renewcommand{\arraystretch}{1.15}
\begin{tabularx}{\columnwidth}{@{}>{\raggedright\arraybackslash}p{0.24\columnwidth} c >{\raggedright\arraybackslash}X@{}}
\toprule
\textbf{Group} & \textbf{N} & \textbf{Tools} \\
\midrule
Session \& state & 3 & \nolinkurl{get_capabilities}, \nolinkurl{get_microscope_state}, \nolinkurl{set_microscope_mode} \\
Acquisition & 5 & \nolinkurl{acquire_tem_image}, \nolinkurl{acquire_stem_image}, \nolinkurl{acquire_4dstem}, \nolinkurl{acquire_eels}, \nolinkurl{acquire_diffraction} \\
Stage & 3 & \nolinkurl{get_stage_position}, \nolinkurl{set_stage_position}, \nolinkurl{run_tilt_series} \\
Optics \& beam & 6 & \nolinkurl{get_optics_state}, \nolinkurl{set_focus}, \nolinkurl{set_stigmation}, \nolinkurl{set_beam_shift}, \nolinkurl{set_beam_blanker}, \nolinkurl{run_beam_alignment} \\
Detectors & 3 & \nolinkurl{list_detectors}, \nolinkurl{insert_detector}, \nolinkurl{retract_detector} \\
Live-job lifecycle & 4 & \nolinkurl{start_live_processing_job}, \nolinkurl{get_live_processing_job_status}, \nolinkurl{get_live_processing_job_result}, \nolinkurl{stop_live_processing_job} \\
Analysis & 6 & \nolinkurl{compute_virtual_image}, \nolinkurl{compute_com_dpc}, \nolinkurl{compute_radial_profile}, \nolinkurl{compute_fft}, \nolinkurl{find_peaks}, \nolinkurl{match_dspacing} \\
\midrule
\textbf{Total} & \textbf{30} & \\
\bottomrule
\end{tabularx}
\end{table}
\section{Implementing adapters for vendor platforms}\label{app:vendors}
This appendix gives implementation guidance for adapting the \adapter{} contract (Fig.~\ref{fig:adapter-contract}) to vendor-specific platforms. The aim is to make the bridge pattern reproducible without binding the published method to any one vendor SDK; complete adapter implementations and live-instrument validation are reported separately under the corresponding intellectual-property disclosure \citep{dosReis2025Disclosure}.

\subsection{Categorising the host binding}\label{app:host-binding}

Vendor scripting surfaces fall into three categories with respect to the host-process binding friction of Section~\ref{sec:intro}. Table~\ref{tab:vendor-comparison} summarises the categorisation against four representative platforms.

\paragraph{In-process Python.} The vendor exposes a Python module that loads only inside the running acquisition application (e.g.\ Gatan GMS, Thermo Fisher Velox). The MCP server cannot import the module directly; the adapter requires a bridge plugin that runs as a script inside the vendor application and speaks the \texttt{nuance-mcp-bridge/1.0} contract over loopback.

\paragraph{Socket-based service.} The vendor provides a Python client library (e.g.\ JEOL pyJEM against TEMServer) that connects to a long-running service over a TCP socket. The MCP server can either import the client library directly when it runs on the instrument PC, or relay through a ZeroMQ proxy when it runs on a separate workstation.

\paragraph{COM/.NET automation.} The vendor exposes a COM or .NET automation interface with no native Python binding (e.g.\ Hitachi ExTOPE). The adapter requires both a process crossing and a language crossing; the recommended pattern is a small .NET service that wraps the automation objects and speaks the bridge contract over a ZeroMQ socket.

\begin{table*}
\caption{Categorization of representative vendor scripting platforms
against the host-process binding friction of
Section~\ref{sec:intro}. Each row identifies the host-binding
category, the scripting surface, and the resulting bridge strategy.
The list is illustrative rather than exhaustive; instruments from
the same vendor across product lines may use different surfaces.}
\label{tab:vendor-comparison}
\centering
\small
\begin{tabularx}{\textwidth}{@{}>{\raggedright\arraybackslash}X >{\raggedright\arraybackslash}X >{\raggedright\arraybackslash}X c >{\raggedright\arraybackslash}X@{}}
\toprule
\textbf{Vendor (representative platform)} & \textbf{Scripting surface} & \textbf{Host binding category} & \textbf{Process crossing} & \textbf{Bridge strategy} \\
\midrule
Gatan (GMS 3.x, DigitalMicrograph)                            & Python (in-process)   & In-process Python      & Required & ZeroMQ plugin inside the vendor process \\[4pt]
JEOL (TEMServer + pyJEM)                                      & Python (socket-based) & Socket-based service   & Optional & Direct client import or ZeroMQ relay \\[4pt]
Hitachi (ExTOPE / SU-Series SDK)                              & COM / .NET            & COM/.NET automation    & Required & .NET service $\to$ ZeroMQ bridge \\[4pt]
Thermo Fisher Scientific (Velox)                              & Python (in-process)   & In-process Python      & Required & ZeroMQ plugin inside the vendor process \\
\bottomrule
\end{tabularx}
\end{table*}

\subsection{Bridge plugin skeleton}\label{app:bridge-skeleton}

For host-bound platforms (in-process Python and COM/.NET automation), the bridge plugin follows a common pattern, sketched here in a vendor-neutral form. The plugin (i)~binds a ZeroMQ \verb|REP| socket to loopback, (ii)~enters a poll-and-dispatch loop, (iii)~pumps the host application's event loop between polls so the GUI remains responsive, and (iv)~translates each incoming bridge request into a call against the vendor SDK. Listing~\ref{lst:bridge} shows the structure.

\begin{lstlisting}[caption={Vendor-neutral bridge plugin skeleton.
The plugin runs inside the vendor application process and translates
\texttt{nuance-mcp-bridge/1.0} requests into vendor-SDK calls. The
\texttt{vendor\_pump\_events()} placeholder corresponds to a
vendor-specific call (e.g.\ \texttt{DM.PumpEvents()} for GMS). The
\texttt{vendor\_dispatch()} function maps method names defined in
the \adapter{} contract to the vendor SDK's corresponding
operations.},
label={lst:bridge}]
# bridge_plugin.py (runs inside the vendor host process)
import zmq, json, base64
# import vendor SDK here, e.g. DigitalMicrograph

ctx = zmq.Context()
sock = ctx.socket(zmq.REP)
sock.bind("tcp://127.0.0.1:5555")

while True:
    vendor_pump_events()  # keep host GUI responsive
    if sock.poll(100):    # 100 ms poll cadence
        msg = json.loads(sock.recv())
        # Negotiate on first call:
        if msg["method"] == "hello":
            sock.send(json.dumps({
                "status": "ok",
                "result": {
                    "version": "nuance-mcp-bridge/1.0",
                    "capabilities": ADAPTER_CAPABILITIES,
                },
            }).encode())
            continue
        result = vendor_dispatch(msg["method"], msg["params"])
        sock.send(json.dumps({
            "status": "ok", "result": result
        }).encode())
\end{lstlisting}

For socket-based vendor services (e.g.\ JEOL pyJEM against TEMServer), the bridge plugin is unnecessary: the Python-side \adapter{} subclass calls the vendor client library directly. The same listing applies if the MCP server runs on a separate workstation and the vendor client library is imported through a ZeroMQ relay.

\subsection{Implementing the Python-side adapter}\label{app:adapter-impl}

On the server side, each vendor adapter is a subclass of \adapter{}. The subclass (i)~declares the capability set the bound instrument actually supports, (ii)~implements only those methods, and (iii)~forwards each call either through the bridge socket (for host-bound platforms) or directly through the vendor client library (for socket-based platforms). Listing~\ref{lst:adapter} sketches the pattern, with the bridge-roundtrip helper factored out so that adapters for either host-binding category share the bulk of their implementation.

\begin{lstlisting}[caption={Vendor-neutral adapter pattern. The
subclass declares the capability set it supports and implements
the corresponding \adapter{} methods. The
\texttt{\_bridge\_call} helper hides the loopback round trip; for
socket-based vendor services, the same helper can wrap the vendor
client library directly. Schema validation occurs in the FastMCP
server before any call reaches this layer
(Section~\ref{sec:schema}).},
label={lst:adapter}]
# vendor_adapter.py
import zmq, json
from nuance_mcp.adapter import MicroscopeAdapter

class VendorAdapter(MicroscopeAdapter):
    capabilities = {
        "tem", "stem", "stage", "stage.tilt",
        "optics", "detectors",
        # ... whatever the bound instrument actually supports
    }

    def __init__(self, endpoint="tcp://127.0.0.1:5555"):
        ctx = zmq.Context()
        self._sock = ctx.socket(zmq.REQ)
        self._sock.connect(endpoint)
        # Handshake on connect:
        self._bridge_call("hello", {})

    def _bridge_call(self, method, params):
        self._sock.send(json.dumps(
            {"v": "nuance-mcp-bridge/1.0",
             "method": method,
             "params": params}).encode())
        return json.loads(self._sock.recv())["result"]

    async def set_stage_position(self, x_um=None, y_um=None,
                                 z_um=None, alpha_deg=None,
                                 beta_deg=None):
        return self._bridge_call("set_stage_position", {
            "x_um": x_um, "y_um": y_um, "z_um": z_um,
            "alpha_deg": alpha_deg, "beta_deg": beta_deg,
        })

    async def acquire_tem_image(self, exposure_s=1.0,
                                binning=1, roi=None):
        return self._bridge_call("acquire_tem_image", {
            "exposure_s": exposure_s,
            "binning": binning, "roi": roi,
        })
\end{lstlisting}

\subsection{Declaring capabilities and testing the adapter}\label{app:adapter-test}

A new adapter declares only the capability families it actually supports; the default implementations inherited from \adapter{} raise \texttt{CapabilityUnavailable} for the rest, which the tool layer surfaces to the agent as a structured \verb|{"status": "UNSUPPORTED"}| response (Section~\ref{sec:adapter}). Skills that require an undeclared capability fail at their opening \tool{get\_capabilities} check rather than mid-sequence.

The hardware-independent test suite is reusable across adapters: it exercises the schema layer, the dispatch path, the live-job lifecycle, and the bridge JSON contract through the \adapter{} interface and does not import any vendor SDK. A new adapter can therefore be developed against the reference simulator first (which shares the schema layer), then connected to the bridge, and finally validated on the live instrument once the bridge round trip is observed.


\end{document}